\documentclass[12pt]{article}
\usepackage{epsfig}
\newcommand{\be}{\begin{equation}} 
\newcommand{\ee}{\end{equation}}
\newcommand{\bea}{\begin{eqnarray}}
\newcommand{\eea}{\end{eqnarray}}
\begin{document}
\begin{flushright} 
SINP/TNP/03-25\\
\end{flushright}         
\begin{center}
   \vskip 3em
{\Large\bf Quantum Inverse Square Interaction}
\footnote
{Talk presented at the conference
\lq\lq Space-time and Fundamental Interactions: Quantum Aspects''
in honour of Prof. A.P.~Balachandran's 65th birthday, Vietri sul Mare, 
Italy, 26 - 31 May, 2003.}
  
\vskip 2em
\end{center}
\begin{center}
{\large Kumar. S. Gupta}
\vskip 1em
Theory Division,\\
 Saha Institute of Nuclear Physics,\\
1/AF Bidhannagar, Calcutta - 700064, India.
\end{center}
\vskip 1em
\begin{abstract}
Hamiltonians with inverse square interaction potential occur in the study of
a variety of physical systems and exhibit a rich  mathematical
structure. In this talk we briefly mention some of the applications of such
Hamiltonians and then 
analyze the case of the N-body rational Calogero model as an example.
This model has recently been shown to admit novel solutions, whose   
properties are discussed.
\end{abstract}

\setcounter{footnote}{0}

\newpage

\def\be{\begin{equation}} 
\def\ee{\end{equation}}   
\def\bea{\begin{eqnarray}}
\def\eea{\end{eqnarray}}

\section{Introduction}

	Hamiltonians with inverse square type interaction potential appear
in the study of a variety of systems and exhibit rich physical and
mathematical structures. One of the first applications of such Hamiltonians
occurred in the study of electron capture by polar molecules with permanent
dipole moments \cite{fermi}. It has been found that only molecules with a
critical value of dipole moment can capture electrons to form stable anions.
Various proposals have been put forward to explain such a critical value 
\cite{levy,cam} and 
extensive theoretical and experimental studies of such molecular systems are
found in the literature \cite{mol}. It is however a puzzling fact that 
while most theories of such molecules predict the existence of 
an infinite number of bound states, experiments so far have typically detected 
only a single bound state in such systems \cite{mol}.

	Physics of black holes is another interesting area where
Hamiltonians with inverse square type interactions occur \cite{town}. 
These operators appear naturally in the analysis of the near-horizon
properties of black holes \cite{trg} and their study provides an alternate
derivation \cite{ssen} of the near-horizon conformal structure of black
holes \cite{carlip}. The Bekenstein-Hawking entropy of black holes can also
be understood within this framework \cite{ssen} and recently there has been
a novel proposal to understand black hole decay using this formalism
\cite{decay}.

	On a more formal side, Hamiltonians with inverse square interactions 
have been studied within the context of conformal quantum mechanics \cite{fub}.
The Hamiltonian appears as one of the generators of $SU(1,1)$ which plays
the role of the spectrum generating algebra. 
Such systems provide a quantum mechanical setting to analyze the phenomena of 
scaling anomalies \cite{cam,cam1} and renormalization \cite{rajeev}.

	Finally, following the seminal works of Calogero \cite{calo,calo3},
exactly solvable many body systems with inverse square interaction have been
studies extensively in the literature \cite{pr}.
The Calogero model and its variants are 
relevant to the study of many branches of contemporary physics, including
generalized exclusion statistics \cite{poly},
quantum hall effect \cite{qhe}, Tomonaga-Luttinger liquid \cite{ll}, quantum  
chaos \cite{rmt},  quantum electric transport in mesoscopic system
\cite{qet},
spin-chain models \cite{hs} and Seiberg-Witten theory \cite{sw} 
The spectrum of the $N$-particle rational Calogero model was first obtained
almost three decades ago \cite{calo3}, which has since
been analyzed using a variety of different techniques \cite{brink}. 
In the rest of this talk, we shall
describe some new solutions to the N-body rational Calogero model 
that have been found recently \cite{us}. These solutions are qualitatively
very different compared to what was obtained by Calogero.
We shall also see that these new solutions
captures many of the interesting physical and mathematical
aspects of Hamiltonians with inverse square interactions.

\newpage

\section{N-Body Rational Calogero Model}

The rational Calogero model is described by $N$ identical particles
interacting
with each other through a long-range inverse-square and harmonic
interaction on the line.
The Hamiltonian of this model is given by \cite{calo3}
\be
H = - \sum^{N}_{i=1} \frac{{\partial}^2}{\partial x_i^2} +
\sum_{i \neq j} \left [ \frac{a^2 - \frac{1}{4}}{(x_i - x_j)^2} +
\frac{\Omega^2}{16} (x_i - x_j)^2 \right ]
\label{e0}
\ee
where $a$, $\Omega$ are constants, 
$x_i$ is the coordinate of the $i^{\rm th}$ particle and
units have been chosen such that $2 m {\hbar}^{- 2} = 1$.
We are interested in finding normalizable solutions of the
eigenvalue problem
\be
H \psi = E \psi.
\label{e1}
\ee
Following Calogero \cite{calo3}, we consider the above eigenvalue equation
in a sector of configuration
space corresponding to a definite ordering of particles given by
$x_1 \geq x_2 \geq
\cdots \geq x_N$. The translation-invariant
 eigenfunctions of the Hamiltonian $H$ can be written as
\be
\psi = \prod_{i <j} \left (x_i - x_j \right )^{a + \frac{1}{2}} \
\phi (r) \ P_k (x),
\label{e2}
\ee  
where $x \equiv (x_1, x_2, \dots, x_N)$, $r^2 = 
\frac{1}{N} \sum_{i < j} (x_i - x_j)^2$ and $P_k (x)$ is a 
translation-invariant as well as  homogeneous 
polynomial of degree $k(\geq 0)$ which satisfies the
equation
\be
\left[ \sum^{N}_{i=1}\frac{{\partial}^2}{\partial x_i^2}
+ \sum_{i \neq j} \frac{ 2 (a + \frac{1}{2}) }{(x_i -
x_j)}  \frac{{\partial}}{\partial x_i} 
\right] P_k (x) = 0.
\label{e4}
\ee
The existence of 
complete solutions of (4) has been discussed by Calogero
\cite{calo3}.
Substituting Eqn. (\ref{e2}) in Eqn. (\ref{e1}) and using Eqns (4) we get
\be
\tilde{H} \phi
= E \phi,
\label{e5}
\ee  
where
\be
\tilde{H} = \left [ - \frac{d^2}{dr^2} - (1 + 2 \nu )
\frac{1}{r} \frac{d}{d r} + w^2 r^2 \right ] 
\label{e9}
\ee
with $w^2 = \frac{1}{8} \Omega^2 N $ and
\be  
\nu = k + \frac{1}{2}(N - 3) + \frac{1}{2} N (N-1)(a + \frac{1}{2}).
\label{e6}
\ee
$\tilde{H}$ is the effective Hamiltonian in the ``radial'' direction.
It can be shown that $\phi(r) \in L^2[R^+,d\mu]$
where the measure is given by $d\mu = r^{1 + 2 \nu} dr$.

The Hamiltonian $\tilde{H}$ is a symmetric (Hermitian) operator on the domain
$D(\tilde{H}) \equiv \{\phi (0) = \phi^{\prime} (0) = 0,~
\phi,~ \phi^{\prime}~  {\rm absolutely~ continuous} \} $. 
To determine whether 
$\tilde{H}$ is self-adjoint \cite{reed} in $D(\tilde{H})$, 
we have to first look for square integrable solutions of the equations 
\be
\tilde{H^*} \phi_{\pm} = \pm i \phi_{\pm},
\label{e10}
\ee
where $\tilde{H^*}$ is the adjoint of $\tilde{H}$ (note that $\tilde{H^*}$
is given by the same differential operator as $\tilde{H}$ although their
domains might be different). 
Let $n_+(n_-)$ be the total number of square-integrable, independent solutions 
of (\ref{e10})
with the upper (lower) sign in the right hand side. Now $\tilde{H}$ 
falls in one of the following categories \cite{reed} : 
1) $\tilde{H}$ is (essentially) self-adjoint iff
$( n_+ , n_- ) = (0,0)$;
2) $\tilde{H}$ has self-adjoint extensions iff $n_+ = n_- \neq 0$;
3) If $n_+ \neq n_-$, then $\tilde{H}$ has no self-adjoint extensions.

For the Calogero model, the solutions of Eqn. (8) are given by
\be
\phi_{\pm} (r) = {\mathrm e}^{- \frac{w r^2}{2}} 
U \left ( d_\pm, c, w r^2 \right ),
\label{e11}
\ee
where $d_{\pm} = \frac{1+ \nu }{2} \mp \frac{i}{4 w}$, $c = 1+ \nu $ and  $U$
denotes the confluent hypergeometric function of the
second kind \cite{abr}. The asymptotic behaviour of $U$ \cite{abr} together
with the exponential factor in Eqn. (9) ensures that 
$\phi_{\pm} (r)$ vanish at infinity. The solution in Eqn. (9) have
different short distance behaviour for $\nu \neq 0$ and $\nu = 0$. 
From now onwards, we shall restrict our discussion to the case for 
$\nu \neq 0$, the analysis for $\nu = 0$ being similar. When $\nu \neq 0$, 
$U(d_{\pm},c,w r^2)$ can be written as 
\be
U \left ( d_{\pm},c,w r^2 \right )=
C
\bigg [ \frac{M \left ( d_{\pm}, c, w r^2 \right )}
{\Gamma (b_{\pm}) \Gamma (c)}
 -  \left ( w r^2 \right )^{1 -c}
\frac{M \left ( b_{\pm}, 2-c, w r^2 \right )}
{\Gamma (d_{\pm}) \Gamma (2-c)} \bigg ],
\label{e12}
\ee
where $b_\pm = \frac{1-\nu}{2} \mp \frac{i}{4w}$, $C = \frac{\pi}{{\mathrm
sin} (\pi + \nu \pi)}$ and 
$M$ denotes the confluent hypergeometric function of the first kind
\cite{abr}.
In the limit $r \rightarrow 0$, $M(d_{\pm},c,wr^2) \rightarrow 1$. 
This together with Eqns. (9) and (10) implies that as
$r \rightarrow 0$,
\be
|\phi_{\pm} (r)|^2 d \mu \rightarrow 
\left [ A_1 r^{(1 + 2 \nu )} + A_2 r + A_3 r^{(1 - 2 \nu )} \right ] dr,
\label{e14}
\ee
where $A_1, A_2$ and $A_3$ are constants independent of $r$. 
From Eqn. (11) it is
now clear that in the limit $r \rightarrow 0$, the functions
$\phi_{\pm} (r)$ are not square-integrable if ${\mid \nu \mid} \geq 1 $.
In that case, $n_+ = n_- = 0$ and $\tilde{H}$ is essentially self-adjoint
in the domain $D(\tilde{H})$. However, if either $ 0 < \nu < 1$ or
$ -1 < \nu < 0$, the functions $\phi_{\pm} (r)$ are indeed square-integrable.
Thus if $\nu$ lies in these ranges, we have $n_+ = n_- = 1$ and 
Hamiltonian $\tilde{H}$ is not self-adjoint in $D(\tilde{H})$ but
admits self-adjoint extensions.
The domain $D_z(\tilde{H})$ in which $\tilde{H}$ is self-adjoint contains all
the elements of $D(\tilde{H})$ together with elements of the form
$\phi_+ + {\mathrm e}^{iz} \phi_-$, where $ z \in R$ (mod $2 \pi$) \cite{reed}.
We can similarly show that $n_+ = n_- = 1$ for $\nu = 0$ as well. Thus the
self-adjoint extensions of this model exist when $-1 < \nu < 1$.

In order to determine the spectrum 
we note that the solution to Eqn. (5) which is
bounded at infinity is given by 
\be
\phi (r) = B {\mathrm e}^{- \frac{w r^2}{2}} 
U(d,c,w r^2),
\label{e15}
\ee
where $d = \frac{ 1+ \nu }{2} - \frac{E}{4 w}$ and $B$ is a constant.
In the limit $r \rightarrow 0$, 
\be
\phi (r) \rightarrow 
B C \left [ \frac{1}{\Gamma (b) \Gamma (c)}  
- \frac{w ^{-\nu}  r^{-2 \nu }}{\Gamma (d) \Gamma (2-c)} \right ],
\label{e16}
\ee
where $b = \frac{1 - \nu}{2} - \frac{E}{4w}$.
On the other hand, as $r \rightarrow 0$,
\be
\phi_+ + {\mathrm e}^{iz} \phi_-  \rightarrow  C \bigg [
\frac{1}{\Gamma (c)}  \left (
\frac{1}{\Gamma (b_+) }
+\frac{{\mathrm e}^{iz}}{\Gamma (b_- ) } \right )
  -  \frac{ w ^{-\nu} r^{-2 \nu }}{\Gamma (2-c)}   
\left ( \frac{1}{\Gamma (d_+) } 
+\frac{{\mathrm e}^{iz}}{\Gamma (d_-) } \right ) \bigg ].
\label{e17}
\ee
The requirement of self-adjointness demands that 
$\phi (r) \in D_z(\tilde{H})$, which implies that the coefficients of 
different powers of
$r$ in Eqns. (13) and (14) must match. Comparing the coefficients 
of the constant term and $r^{-2 \nu }$ in 
Eqns. (13) and (14) we get
\be
f(E) \equiv \frac{\Gamma \left ( \frac{ 1 - \nu }{2} - \frac{E}{4 w} \right )}
{\Gamma \left (\frac{ 1 + \nu }{2} - \frac{E}{4 w} \right ) } =
\frac{\xi_2 {\mathrm cos}(\frac{z}{2} - \eta_1)}
{\xi_1 {\mathrm cos}(\frac{z}{2} - \eta_2)},
\label{e18}
\ee 
where $\Gamma \left ( \frac{ 1 + \nu }{2} + \frac{i}{4 w} \right )
\equiv \xi_1 {\mathrm e}^{i \eta_1}$
and
$\Gamma \left ( \frac{ 1 - \nu }{2} + \frac{i}{4 w} \right )
\equiv \xi_2 {\mathrm e}^{i \eta_2}$.
For given values of the parameters $\nu$ and  $w$, 
the bound state energy $E$ is obtained from Eqn. (15) as a function of
$z$. The corresponding eigenfunctions are
obtained by substituting $\phi(r)$ from Eqn. (12) into Eqn. (3).
Different choices of $z$ thus leads to 
inequivalent quantizations of the many-body Calogero model.
Moreover from Eqn. (\ref{e18}) we see that for fixed value of $z$, 
the Calogero model with 
parameters $(w, \nu)$ and $(w,-\nu)$ produces identical energy 
spectrum although the corresponding wavefunctions are different.


\begin{figure}
\begin{center}
\includegraphics[width=0.6\textwidth, height=0.3\textheight]{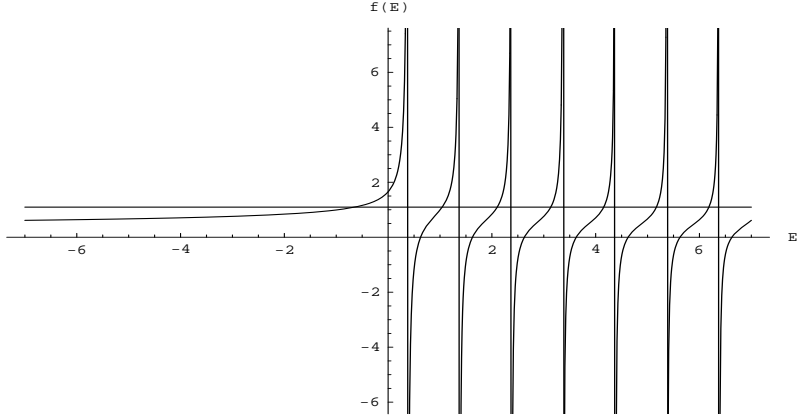}
\caption { \label{fig1} A plot of Eqn. (\ref{e18}) using Mathematica
with $w = 0.25$, $\nu = 0.25 $ and
$z = -1.5$. The horizontal straight line corresponds the value of the r.h.s
of
Eqn. (\ref{e18}).}
\end{center}
\end{figure}

\newpage
The following features about the spectrum may be noted: \\

\noindent
1) We have obtained the spectrum analytically when the r.h.s.
of Eqn. (15)
is either 0 or $\infty$. When the
r.h.s. of Eqn. (15) is $0$, we must have the situation where 
$\Gamma \left (\frac{ 1 + \nu }{2} - \frac{E}{4 w} \right )$ blows up, i.e.
$E_n = 2 w ( 2 n + \nu + 1)$ where $n$ is a positive integer. 
This happens for the special choice of $z = z_1 = \pi + 2 \eta_1$.
These
eigenvalues and the corresponding eigenfunctions are analogous to those
found by Calogero although for a different parameter range. 
Similarly, when the r.h.s. of Eqn. (15) is $\infty$, 
an analysis similar
to the one above shows that $E_n  = 2 w ( 2 n - \nu + 1)$. 
This happens for the special value of $z$ given by 
$z = z_2= \pi + 2 \eta_2$. \\

\noindent
2) For choices of $z$ other than $z_1$ or $z_2$, the nature of the spectrum
can be understood from Figure 1, which is a plot of Eqn. (15) for
specific values of $\nu, z$ and $w$.
In that plot, the horizontal straight line  corresponds to the 
r.h.s of Eqn. (15). The energy eigenvalues are obtained from the
intersection of $f(E)$ with the horizontal straight line.
Note that the spectrum
generically consists of infinite number of positive energy solutions and at
most one negative energy solution. The existence of the negative energy
states can be understood in the following way. For large negative values of
$E$, the asymptotic value of $f(E)$ is given by $(\frac{E}{4w})^{- \nu}$
\cite{abr}, which monotonically tends to 0 or $+ \infty$ 
for $\nu > 0$ or $\nu < 0$ respectively. When $\nu > 0$, the negative energy
state will exist provided r.h.s. of Eqn. (15) lies between 0 and 
$\frac{\Gamma  ( \frac{ 1 - \nu }{2})}
{\Gamma  (\frac{ 1 + \nu }{2} )}$. Similarly, when $\nu < 0$, the 
negative energy state will exist when the r.h.s. of Eqn. (15)
lies between 
$\frac{\Gamma  ( \frac{ 1 - \nu }{2})}
{\Gamma (\frac{ 1 + \nu }{2} ) }$ and $+ \infty$.
For any given values of $\nu$ and $w$, the 
position of the horizontal straight line in Fig. 1 can always be adjusted 
to lie anywhere between $-\infty$ and $+\infty$ by 
suitable choices of $z$. Thus the spectrum would always contain a negative
energy state for some choice of the parameter $z$.    \\

\noindent
3) Contrary to the spectrum of the rational Calogero model, 
the energy spectrum obtained from Eqn. (15) is not equispaced for
finite
values of $E$ and for generic values of $z$. 
This may seem surprising with the presence of $SU(1,1)$ as
the spectrum generating algebra in this system \cite{fub}, which demands 
that the eigenvalues be evenly spaced. In order to address this issue, we   
consider the action of the dilatation generator $D = \frac{1}{2} \left ( r  
\frac{d}{dr} + \frac{d}{dr}r \right )$ on an element 
$\phi(r) = \phi_+(r) + {\mathrm e}^{iz} \phi_-(r)$.  
In the limit $r \rightarrow 0$, we have
\be
D \phi = \frac{C}{2} \bigg [
\frac{1}
{\Gamma (c)} \left (
\frac{1}{\Gamma (b_+) }
+\frac{{\mathrm e}^{iz}}{\Gamma (b_-) } \right )
  - \frac{r^{-2 \nu } (1 - 4 \nu) }
{\Gamma (2-c)}
\left ( \frac{1}{\Gamma (d_+) }
+\frac{{\mathrm e}^{iz}}{\Gamma (d_-) } \right ) \bigg ].
\label{e19}
\ee
We therefore see that $D \phi (r) \in D_z (\tilde{H})$ only for $z = z_1$ or
$z= z_2$.
Thus the generator of dilatations does not in
general leave the domain of the Hamiltonian invariant
\cite{dh,us}. Consequently, $SU(1,1)$ cannot be implemented as
the spectrum generating algebra except for $z = z_1$, $z_2$.\\

We have thus presented a new quantization scheme for the rational
Calogero model. The non-equispaced nature of the energy levels and the 
existence of a negative energy bound state are some of the salient features
that emerge from our analysis. These features of the solutions presented
here are qualitatively different from those obtained by Calogero.

\section{Conclusion}

In this talk we have mentioned some of the interesting areas in which 
Hamiltonian operators with inverse square interaction occur. In some of the
applications that we have analyzed, including the Calogero model discussed
above, the self-adjoint extension of the relevant Hamiltonian operator
plays an important role. Such Hamiltonians are also generically related to
the Virasoro algebra, which hints at the existence of an underlying 
conformal symmetry. While the presence of the conformal symmetry has been 
established in some of the examples, it is not clear if such a symmetry
plays any role in a generic case. Much more work needs to be done to
understand the role of such a symmetry, which is a task for the future. 

\section*{Acknowledgments}

The importance of self-adjoint extension in physics has often been
emphasized by Professor A. P. Balachandran which has served as an
inspiration to me in pursuit of this line of research. I am very grateful to
Bal for this. The material presented here is based partly on work done in 
collaboration with
B. Basu-Mallick, D. Birmingham, P. K. Ghosh and S. Sen. I would like to
thank the organizers of the conference in general, and F. Lizzi and G. Marmo
in particular, for their hospitality in Vietri. I also thank the
Associateship Scheme of the Abdus Salam ICTP, Trieste, Italy for their
financial support which made my participation in the conference possible.

\end{document}